\documentstyle[12pt]{article}

\begin{document}

\renewcommand{\theequation}{\arabic{equation}}
\renewcommand{\thesubsection}{\Roman{subsection}}

\def\beqra{\begin{eqnarray}} \def\eeqra{\end{eqnarray}}
\def\beqast{\begin{eqnarray*}} \def\eeqast{\end{eqnarray*}}
\def\beq{\begin{equation}}	\def\eeq{\end{equation}}
\def\be{\begin{enumerate}}   \def\ee{\end{enumerate}}
\def\Om{\Omega}\def\ul{\underline}
\def\nd{\noindent}
\def\haf{\frac{1}{2}}
\def\La{\Lambda}

\vspace*{24pt}
\begin{center}
 \large{\bf IMPROVED BOUNDS ON NON-LUMINOUS \\[6pt]

MATTER IN SOLAR ORBIT}
\normalsize

\vspace{40pt}
John D. Anderson, Eunice  L. Lau,  and Timothy P. Krisher \\

\vspace{12pt}
{\it Jet Propulsion Laboratory \\
California Institute of Technology,
Pasadena, California 91109}

\vspace{24pt}
Duane A. Dicus \\

\vspace{12pt}
{\it Center for Particle Physics and Department of Physics\\
University of Texas, Austin, Texas 78712}

\vspace{24pt}
Doris C. Rosenbaum and Vigdor L. Teplitz \\

\vspace{12pt}
{\it Physics Department  \\
Southern Methodist University,  Dallas, Texas 75275}

\vspace{18pt}
ABSTRACT
\end{center}

We improve, using a larger set of observations including Voyager 2 Neptune
flyby  data,  previous  bounds on the amount of dark matter (DM)
trapped in a spherically symmetric distribution about the sun.
We bound DM by noting that such a distribution would increase the effective
mass of the sun as seen by the outer planets and by finding the uncertainty
in that effective mass for Uranus and Neptune in fits to the JPL
Developmental Ephermeris residuals, including optical data and those two
planets'
Voyager 2 flybys.  We extend our previous procedure by fitting  more
parameters of the developmental ephemerides.  Additionally, we present
here the values for Pioneer 10 and 11 and Voyager 1 and 2 Jupiter ranging
normal points (and incorporate these data as well).  Our principal result is
to limit DM in spherically symmetric distributions in orbit about the sun
interior to Neptune's orbit to less than an earth mass and interior to
Uranus' orbit to about 1/6 of an earth's mass.

\pagebreak
\baselineskip=20pt
\setcounter{page}{1}

\renewcommand{\thesubsubsection}{A.}
\subsection{Introduction}

 \subsubsection{Background}

\vspace{.2cm}
\indent\indent
The purpose of this work is to use Deep Space Network (DSN) radio tracking
data from the Voyager Neptune and  Uranus encounters to investigate limits
on the amount of non-luminous solar halo matter in a spherically symmetric
distribution.  In an earlier work (Anderson et al. 1989) we analyzed
tracking  data  during the Voyager 1986 Uranus encounter.  That encounter
permitted reduction of the $1\sigma$ uncertainty in Uranus' range (at the
encounter time) from about1500km to 1km.  This new constraint on the orbit of
Uranus led to a bound of $3\times 10^{-6}M_\odot$ on the amount of
spherically symmetric, non-luminous matter in solar orbit interior to the
radius of Uranus's orbit, an improvement of at least an order of magnitude
from the bound without the Voyager ranging data.  After that work was
completed, we  continued the analysis by adding more data (optical and
radio from various sources) and by adding the ranging normal point from
the Voyager 2 flyby of Neptune.  This paper presents the improvements in
bounds on spherically symmetric, non-luminous matter that follow from
these new data.

\vspace{.4cm}

\renewcommand{\thesubsubsection}{B.}

 \subsubsection{Method of Analysis}

\vspace{.2cm}
\indent\indent
The basic idea is to compare the effective solar mass felt by the inner
planets to the effective solar masses felt by Uranus and Neptune.  If there is
a spherically symmetric distribution of unseen matter not included in
ephemerides fitting programs, then, when the effective solar mass,
$M_{eff}$, is considered a free parameter for a planet, the value determined
for $M_{eff}$ should be sensitive to the matter interior to its orbit not
otherwise included in the fitting program.  The bound on the difference
between $M_{eff}$ as determined by this method from the motion of an outer
planet and $M_{\odot}$  then constitutes a bound on the total
mass in a spherically symmetric distribution between the inner planets and
that outer planet.  $M_{\odot}$ is determined from the fit to the entire Solar
System in which the value of $M$ is driven by the much more accurately and
precisely known motions of the inner planets.

While one may  determine the collection of
solar system ephemerides, with $M_{\odot}$ different for each planet, we first
adopted a more modest approach.  We determined a value of $M_{eff}$ from
fitting all solar system ephemerides without provision for a varying
$M_{eff}$, i.e. we used the JPL ephemerides, and  the solar mass
determined by it.  We then found new values for $M_{eff}$ for Uranus and
Neptune by refitting for just their ephemerides and the two $M_{eff}$ values
with a data set consisting of the residuals for the Uranus and Neptune
observations (observed minus computed).  A statistically significant
difference between the $M_{eff}$ value and the $M_{\odot}$ value would
constitute
detection of spherically symmetric non-luminous matter; bounds on the
difference constitute bounds on the mass of such a distribution.  In the
present work we first use the same method and then extend the fitting of
residuals to
include the full set of orbital parameters used in the ephemerides program.

\vspace{.4cm}
\renewcommand{\thesubsubsection}{C.}

 \subsubsection{Motivation}

\indent\indent
A major reason for investigating solar halo dark matter is the general
desirability of observing directly as much as possible about the Solar
System.  Beyond this, there are a wide range of specific reasons for
attempting to detect non-luminous solar halo matter or, failing detection,
to place observational bounds on the amount of such matter.  First, there
are many reasons for believing that dark matter exists.  These include the
cosmological  dark matter problem, the galactic cluster dark matter
problem, the galactic halo dark matter problem, and the short-period comet
question.  A recent concise summary of cosmological dark matter problems
is given by Turner (1991) and further background may be found in Kolb and
Turner (1990); the galactic disk dark matter problem is reviewed by
Bahcall (1984,1992); and the short period comet question has been recently
discussed by Weissman (1990).

The cosmological dark matter problem stems from the relative proximity of
the observed cosmic density to the critical density, sharpened by Guth's
observation (1981) that the apparent isotropy, homogeneity, and flatness
of the universe could be explained by a period of exponential inflation.
One consequence of inflation is that $\Om=\rho/\rho_c$, the ratio of
cosmic mass-density to the critical density, should be one.  Since
$\Om_L$, the ratio for luminous matter, is observed to be of the order of
0.01 and the ratio for baryonic matter is bounded by cosmic
nucleosynthesis constraints at about 0.1, ten times as much, it is useful
to search for signs for non-luminous (dark) matter, both baryonic and
non-baryonic, in as many places as possible.  Observational evidence for
non-luminous matter comes from rotation curves in galaxies of Rubin et al.
(1985), Hoffman et al. (1993) and properties of  clusters of galaxies
including galaxtic motion and hot gas distributions.  See Mulchaey et al.
(1993) for recent results.

These issues may need more than one kind of dark matter for resolution.
Indeed the recent COBE results (Smoot et al., 1992) observing large scale
anisotropies  encourage speculation that there may be both hot and cold
dark mater (DM relativistic and non-relativistic at recombination).  Many
kinds of DM have been conjectured, including ordinary baryonic matter in
non-luminous form, axions, supersymmetric particles, massive neutrinos,
black holes, and more exotic particles.  Most candidates are weakly
interacting in order to explain the lack of luminosity, but cross sections
vary according to other desiderata; Press and Spergel (1985) and Faulkner
and Gilliland (1985), for example, use ``cosmions'' to address
simultaneously the solar neutrino and dark matter problems.  They would
have cross sections about 10$^4$ times weak cross sections and hence
could dissipate and be trapped in solar orbit.  Particle detector
searches, however, have left the cosmion dead, or very nearly so: The
results of Caldwell et al. (1990) ``exclude nearly all of the mass range
possible for cosmions'' -- at least for models in which cosmion-nuclear
cross sections scale roughly as the square of the number of nucleons. Many
believe that the most likely candidate is the Lightest Supersymmetric
Particle (LSP).  Supersymmetry assigns to each ``ordinary'
 particle of integral (half integral) spin a supersymmetric partner of
half integral (integral) spin; there is conservation of the total number
of supersymmetric particles in most models.  The existence of an LSP
should be decided early next century from experiments at the
 Large Hadron Collider if it is  constructed.  Other particle physics
candidates are the axion or a massive neutrino.
 See, for example, Kane (1992).

It may be possible, in some of these cases, that a significant density of
non-luminous matter could condense into a halo about a newly forming star.
The conditions on particle masses and interaction cross sections under
which this would be the case, taking into account gravitational
interaction mechanisms in star formation, have not been worked out in
detail but it is difficult to envision mechanisms that would lead to
capture of significant amounts of weakly-interacting DM particles.  This
is because, without some dissipation mechanism, dark matter cannot
concentrate in the galactic disk, be enhanced in giant molecular clouds or
condense  sufficiently in star formation.  Nevertheless dissipation is not
impossible.  A characteristic feature of at least some superstring models
as noted by Gross et al. (1985) and recently discussed by, for example,
Khloper et al. (1991) and Hodges (1993) would have dark matter composed of
mirror or shadow baryons that only interact with normal baryons
gravitationally but could dissipate by emission of undetected shadow
photons.  While this model appears far from compelling, it has an
interesting history of thought behind it (much of it cited by Khloper et
al., 1991 and Hodges, 1993).  (It is however, in serious disagreement with
cosmological Helium ``observations'').  Were it true, there would appear
to be reasonable  likelihood of some concentration of DM particles in the
disk and in giant molecular clouds and perhaps about the sun.  Indeed,
Khloper et al. cite estimates of $10^{-7}$ to $10^{-6}$ solar masses of
shadow matter being captured by a normal matter star.  There are at least
two other models that permit some dissipation in principle, but in
practice are severely constrained over most of their parameter space:
SIMPs (Strongly Interacting Massive Particles) which are reviewed by
Starkman et al. (1990) and CHAMPs (Charged Massive Particles) limits on
which are given by Gould et al. (1991).

A different DM candidate may have been detected.  Recent reports (Alcock
et al., 1993, Auborg et al., 1993, Udalski et al., 1993) have cited
observations by two different groups of what appears to be ``microlensing'' by
a MACHO (Massive Compact Halo Object)  in the halo of our galaxy of a star in
the Large Magellanic Cloud.  MACHOs, such as brown dwarfs, are an important
baryonic DM candidate, but were they to yield an appreciable fraction of the
closure density they would be in serious contradiction with the lower limit on
cosmological deuterium production (because high baryon density leads to
``complete'' burning of deuterium into
${}^4$He).  In short the DM situation is complex and fluid.

A different motivation for bounding non-luminous material trapped in
solar  orbit is the need for observational limits on solar system
components.  Tremaine (1990) has reviewed the subject of dark matter in
the Solar System.  He discusses techniques for measuring DM,  including
the one  used by Anderson et al. (1989), and lists limits set by each.
Tremaine reviews models which would account for DM being trapped in the
formation of the planetary system.  The planets are believed  to have
been formed from a disk of gas and dust surrounding the Sun.  As the
disk  cooled, non-volatile material condensed into ``planetesimals''
many of which are incorporated in the cores of the giant planets.  DM in
the solar system could be in the form of a spherically symmetric
population of residual planetesimals.   Various  forms of DM, including
such bodies should be absent from the inner solar system because of
gravitational perturbations by Jupiter and the inner planets.  However,
residual baryonic DM may be present in at least two locations.  One is
the generally accepted Oort cloud of perhaps 70 to 100 $\;M_{\oplus}$ at
$r>2\times 10^4$ AU (Oort 1950).  The second location is the
controversial Kuiper belt (Kuiper 1951), perhaps the inner boundary of a
flattened core of comets inside $2\times 10^4$ AU.  It is speculated
that the protoplanetary disk may have extended well beyond  Neptune's
semi-major axis of 30 AU.  It is possible that there is a residual mass,
the Kuiper belt, located in the area $30-45$ AU, or so, and made of
matter that was not depleted in the formation of Neptune.  This
hypothesized comet belt would be in the plane of the ecliptic with total
mass of the order of   $M_{\oplus}$.  The Jupiter-family of comets with
periods of less than 20 years (the so called short-period comets) gives
indirect evidence for such a belt.  It should be noted, that, since
Tremaine's review, several objects, about 1.6 billion kilometers beyond
Neptune have been detected  (Jewitt and Luu, 1993). These may be the
first observations of members of the postulated Kuiper belt.

Even though composed of ordinary matter and therefore having large cross
sections with ordinary planets, such planetesimals would probably not have
been accreted onto the planets in the age of the solar system.  Writing
$$dM/dt =\pi R^2v\,\rho$$
where $\rho\sim M_{DM}/(4/3\, \pi\,r^3$) with $R$ the radius of Uranus,
$r\sim 20$AU, and $v$ Uranus' orbital velocity gives
$$ 1/M_{DM} \times dM/dt \times 4.5\times 10^9 ~{\rm years}~ \sim 10^{-2}
\,. $$
Thus, even ordinary matter (planetesimals) might have survived (although
likely not in spherical distribution inside $10^4$ AU) since solar
 system formation.  This ordinary matter is probably not sufficiently
luminous to be detected with current instruments at the low densities
under consideration, even in the infrared (Backman and Gillett, 1987).

For all these specific theoretical reasons and more generally, as noted
above, because it is of interest to search for any additional existing
matter that might possibly be in the Solar System, it is desirable to use
all available data to detect non-luminous matter in solar orbit or,
failing detection, to put bounds on the magnitude of such matter.

Section II below describes our procedure and presents the results.
Section III contains discussion.  Quantitative work in the paper is
restricted to the cases of spherically symmetric distributions of DM in
solar orbit.  Only a very rough statement is made on the Kuiper belt
question.

\subsection{Limits on Trapped Non-luminous Matter}

\vspace{.4cm}
\renewcommand{\thesubsubsection}{A.}

\subsubsection{Analysis}

\indent\indent
We refer the reader to the detailed discussion in Anderson et al.  (1989)
of  our fitting procedure.  Here we review briefly the essence of the
method and the extensions and improvements incorporated into the current
work.

Reduced to its simplest terms, the planet's position vector is
approximated  by the following two-body expression.
\beq
\vec{r} =a(\cos\;E-e)\hat{P}+a\sqrt{1-e^2}\,\sin\, E\hat{Q}
\eeq
where $a$ and $e$ are the semi-major axis and eccentricity of the Kepler
ellipse, while $\hat P$ and $\hat Q$ are the orthogonal unit vectors in the
orbit
plane with $\hat P$ directed to the perihelion.  The eccentric anomaly $E$ is
related to the time $t$ by,
\beq
E-e\,\sin\,E=u_0+nt
\eeq
where $u_0$ is the mean anomaly at the epoch and the fundamental orbital
angular frequency $n$ is related to $a$ and the central mass $M$ by
$GM=n^2a^3$.

For purposes of gaining insight into what is being measured, we linearize
equation (1) with respect to $a$ and $n$.
\beq
\Delta\vec r=\vec r\;\frac{\Delta a}{a} + t\, \frac{d\vec r}{dt}\;
\frac{\Delta n}{n}\,.
\eeq
It is apparent from equation (3) that $a$ is determined by observations in the
radial direction, while $n$ is determined by observations along the velocity
vector.  The angular frequency $n$, or equivalently the sidereal period
$2\pi/n$, is determined by ground-based astrometric observations of the
planetary motion on the sky.  However astrometric observations provide only
a weak determination of $a$ through the heliocentric parallax.  It is the
ranging data that provide a good determination of $a$.

We recall that the mean orbital radius averaged over time is not the
semi-major axis $a$.  Instead, the time average of $1/r$ is $1/a$.
Therefore for a central mass distribution, the circular velocity $v_c(a)$ at
orbital radius $a$ is just $v_c = na$, a product determined by astrometric
and ranging data.  Our data analysis yields either $v_c(a)$, or equivalently
the effective mass of the Sun $GM_{eff}$ interior to orbital radius $a$.  In
the absence of ranging data over a complete orbital revolution, the two
parameters $a$ and $n$ will be correlated.  The full accuracy of the
ranging data will not map directly into the determination of $GM_{eff}$.
Therefore in setting one-sigma error estimates from the data analysis, we
compute the formal covariance matrix for the  \break
$N_P$ parameter least
squares fit, and then multiply the formal errors by a factor of three.
Of course based on random statistics, we would accept the formal errors as they
stand.  We are reluctant to do so, however, because we are certain systematic
errors exist in the optical observations, especially as introduced through
the optical reference frames used for data reduction.  There may even be
significant dynamical systematic errors introduced by unmodeled sources of
gravitation, the Kuiper belt for example or undetected asteroids and comets.
It is not possible to evaluate the precise influence of these systematic
errors because they are quantitatively unknown.  We therefore make a rather
arbitrary decision to call our format three-sigma errors the realistic
one-sigma errors.

Standish (1993) has pointed out the
difficulty of characterizing hypothetical  gravitational sources, in
particular Planet X, using optical observations. Regarding systematic error,
our concern is that we not claim a smaller error than the optical
observations can deliver.  The limiting accuracy for a meridian circle
observation is about one arcsecond.  From Eq. 3, we conclude that a small
positive change in solar mass will cause the angular planetary position on
the sky to advance linearly with the time.  In the worst case, the fractional
accuracy in solar mass will be limited by,
$$
\sigma(M)/M=\sqrt{3}\,\frac{T}{\pi\, t}\,\sigma(\theta)
$$
where $T$ is the planet's sidereal period, $t$ is the observational time
interval, and $\sigma(\theta)$ equals one arcsecond.  But this   is the
absolute worst case, in the sense that the systematic error exactly mimics
the signal we are measuring.  Over several decades of observations, it is
unlikely we will be that unfortunate.  We expect that the error will be
smaller by some factor $1/\sqrt{N}$, where for white noise $N$ is equal to
the number of observations.  In the final results reported in Table 2, we
are assuming $N=36$ for Uranus, $N=11$ for Neptune, and
$N=250$ for Jupiter.  Note that our three-$\sigma$ criterion for setting
realistic error is most optimistic for Jupiter, but it should be because we
have optical data over six full orbital revolutions.  For Uranus, and
particularly for Neptune, where we have optical data over less than one
orbital period and only one ranging measurement, we are being quite
conservative in our assumptions on the number $N$ of statistically independent
optical observations.

The Uranus and Neptune radial errors in the DE200 ephemeris were relatively
large because of errors in outer-planet masses.  For Uranus the one-$\sigma$
error was 1500 km (Anderson et al., 1989), while for Neptune it was 8700 km.
Using the Voyager flyby mass results, one could reduce the radial errors to
500 km for Uranus and 2600 km for Neptune.  However, the Voyager 2 flyby
determinations of orbital radii are much  more accurate (one-$\sigma$ error
equal to one km).  We recommend the use of these Voyager radii in future
ephemerides.  Note from Table 1 that the actual DE200 radial errors as
determined by Voyager were 147 km (0.1 $\sigma$) for Uranus and 8224 km (0.9
$\sigma$) for Neptune.  With regard to the ranging measurements in our
earlier works (Anderson et al. 1989) we assumed a 500 m accuracy for the
distance determination to Uranus.  After doing a similar analysis of Voyager
data for Neptune, we are confortable with a 1000 m error estimate (one
$\sigma$) for both Uranus and Neptune.  In all analysis in this paper, we
assumed the error estimates given in Table 1.

\vspace{.4cm}
\renewcommand{\thesubsubsection}{B.}

\subsubsection{Astrometric and Ranging Observations}

\indent\indent
Ideally, we would like to have both astrometric and ranging observations
over a complete orbital period.  Given such data, our determination of
each planet's orbital radius $a$ and angular frequency $n$ would be
uncoupled.  However our data are incomplete in two ways.  First, we have a
limited amount of recent outer-planet VLA (Very Large Array)
radio-interferometric data.  Over a longer
time interval dating from 1830, we have less-accurate meridian circle
(transit) observations.  When carefully reduced, these data are accurate
to about 1.2 arcsec before the introduction of the impersonal micrometer
in 1911, and to about 0.4 arcsec after that.  We have used only the post
1911 data in this work.  Consequently we have astrometric data on Uranus
over slightly less than one orbital period, and on Neptune over about
one-half its orbital period.  We have downweighted the
radio-interferometric data by a factor of 1000, effectively removing it
from our fit.

Secondly, our data are incomplete because outer-planet ranging data are
presently available only during spacecraft flybys.  Thus we have only one
range fix on Uranus and Neptune from the respective Voyager 2 flybys.
Doppler and ranging data generated by the DSN (deep space network) with
Voyager 1 and 2 during their outer-planet flybys are archived in the
National Space Science Data Center (NSSDC).  The Pioneer 10 and 11
spacecraft were not equipped with a ranging transponder, but during their
flybys of Jupiter we introduced a ramp into the DSN's radio transmission and
obtained a rough measure of range by autocorrelating the received and
transmitted ramps.  These Pioneer 10 and 11 Doppler data are also archived
in the NSSDC.

Our reductions of all the currently available flyby data yield the
ranging residuals displayed in Table 1.  In the future we expect to
supplement these reduced data with existing DSN Doppler and ranging data
generated during two Voyager flybys of Saturn and one Ulysses flyby of
Jupiter, as well as with anticipated Jupiter data from the two-year
Galileo orbital tour (December  1995 to December 1997), and four years of
Saturn data during the Cassini tour scheduled for the years 2004 to 2008.
However within the next decade, at least, we expect no qualitative
improvements comparable with those of this work in limits on a spherical
DM distribution.

The numerous data sets included in recent JPL ephemerides have been
reviewed by Standish (1990).  These sets include data that were
unavailable in 1980 when JPL constructed the fundamental planetary and
lunar ephemerides (DE200/LE200) for the {\it Astronomical Almanac}
(Standish et al., 1992).  For the analysis summarized here, we used a
1993 reference planetary and lunar ephermeris  DE242, along with its
associated astronomical constants, and determined corrections to the
parameters by the method of weighted least squares.  In our previous
analysis using DE111 (Anderson et al., 1989) we determined corrections to
the orbits of Uranus and Neptune only, along with the effective solar mass
for each planet.  In the current analysis, recognizing that a solution for
only two planets produces an ephermeris that is dynamically inconsistent,
we expanded the parameter set to include all the planets, except Pluto,
and all 194 parameters that went into the construction of
DE242.  Although we doubted that our previous dynamically inconsistent
method would significantly alter our conclusions, we nevertheless obtained
the dynamically consistent solution with little additional effort.

We express residuals with respect to the Astronomical Almanac's
 planetary ephemerides (DE200/LE200) available on magnetic tape for the
period 1600-2200.  We feel it is more useful to refer residuals to the
universally available DE200, rather than the temporary JPL ephemeris
DE242 used in this paper.  The Voyager Jupiter residuals are larger than
Pioneer because DE200, created in 1981, included ranging data from the
Pioneer flybys in 1973 and 1974, but not the Voyager flybys in 1979.
The two Pioneer points were in the fit, the Voyager points were not.

In summary, we used reduced right ascension and declination observations
of all the planets except Pluto.  These included optical meridian
transit observations of the Sun and planets from Washington (USNO)
between 1911 and 1982, and from Herstmonceux between 1957 and 1982, from
Bordeux between 1985 and 1992, from Tokyo between 1986 and 1988,
photoelectric meridian transits from La Palma between 1984 and 1992,
astrolabe observations from seven observatories between 1969 and 1985,
and stellar occulation timings of Uranian rings between 1977 and 1983
and Neptune's disk between 1981 and 1985.

In addition to the optical data, we used reduced radar ranging data for
the inner planets Mercury and Venus, and spacecraft ranging for Mars
from the 1971-1972 Mariner 9 orbiter, 1976-1982 Viking Landers, and 1989
Phobos 2 orbiter.  Lunar laser ranging between 1969 and 1991 were
included implicitly by means of information arrays (least-squares normal
equations; see Press et al. (1992)).  We used reduced ranging data provided by
spacecraft flybys of Mercury by Mariner 10 (1974 and 1975) and of Venus by the
1990 Galileo flyby.

But the crucial flyby data for our dark-matter search were the DSN Doppler and
ranging data generated with Pioneer 10/11 and Voyager 1/2 at the outer
planets.   Because of their importance, and to collect them in one place, we
list in Table 1 the reduced data in two formats.  In the first we express the
ranging data near the flyby time as a geometric coordinate distance
between the center of Earth and center of the planet.  The coordinates for
the geometry are isotropic metric coordinates as described by Standish et
al. (1992).  In the second format, we list the ranging residuals (observed
minus computed) referred to DE200.   An advantage of the residuals
is that they remain essentially constant over the duration of the flybys,
while the geometric distances apply only to the precise  times listed in
Table 1.

\vspace{.4cm}
\renewcommand{\thesubsubsection}{C.}

\subsubsection{Dark Matter Bounds}

\indent\indent
Table 2 gives the results of our fits.
 Line 1 gives the results of Anderson et al., (1989) $M_{DM}(r<r_U) <
2.8\times 10^{-6}M_{\odot}$ and $M_{DM}(r<r_N)<114\times
10^{-6}M_\odot$.  Line 2
shows, for comparison, the improvement that results from the new
fitting procedure used without the Neptune ranging.  Lines 3-5 show the
dramatic effect of  including  the Neptune ranging data:
the bound on $DM$ in spherically
symmetric distribution in orbits interior to the
orbit of Neptune  falls from over $30M_{\oplus}$ to about $M_{\oplus}$.

Lines 3-5 show that the  results are not affected by adding data (or
parameters) for Jupiter.  Note that the minus sign under $DM$ interior
to Neptune's orbit is quite provocative.  If it were statistically
significant, which we cannot claim, one interpretation would be that it
is the effect on Neptune's motion of a non-spherically symmetric mass
distribution exterior to, but relatively close to, Neptune's orbit
(i.e., a Kuiper belt).

\vspace{.4cm}
\renewcommand{\thesubsubsection}{D.}

\subsubsection{The Isothermal Sphere}

\indent\indent
It is tempting to ask for the limit that can be placed on non-luminous
$DM$  under the assumption of a given radial distribution, and the
isothermal sphere is an obvious distribution choice.

For such an analysis we would assume that the distribution of mass in
 the  solar system consists of a centrally condensed source (the Sun)
surrounded by a spherically symmetric dark halo approximated by an
isothermal ideal gas sphere.  Mutual gravitational attraction by the
planets will perturb this configuration, but in a completely
deterministic fashion which could be accounted for in the data
analysis.  At sufficiently large distances from the isothermal core, the
density distribution approaches the power law $r^{-2}$.  Therefore the
effective mass of the Sun at orbital radius $a$ is  \beq
GM(a) = GM_{\odot} + 2 \sigma^2 a
\eeq
where $\sigma$ is the velocity dispersion (rms deviation from the mean)
in  one direction.

$GM_{\odot}$ for the Sun is determined from ranging data for the inner
planets, so  the only unknown in the model of equation (4) is the
velocity dispersion.  Without the isothermal-sphere constraint, we
determine all outer-planet $GM$'s as independent parameters.  As an
alternative, we could impose the constraint given by equation (4) and
refer all $GM$ determinations to the orbital radius $a_7$ of the seventh
planet Uranus.  The linear relation between an arbitrary $GM$ at the
orbital radius $a$ and $GM_7$ for Uranus is
\beq
\Delta\left[GM(a)\right]
=\left(\frac{a}{a_7}\right)\Delta\left[GM(a_7)\right]\,.
\eeq
Therefore, we could impose the isothermal-sphere constraint by
multiplying  the linear coefficient for each $GM$ by $a/a_7$ for Jupiter
and Neptune, and by replacing the three independent $GM$'s by a single
$GM_7$ for Uranus in the least-squares fit.  If we had obtained a
statistically significant determination of $\Delta(GM_7)$, we would have
obtained a determination of the density $\rho$ of dark matter at the
orbital radius of Uranus. \beq
\rho(a_7) = \frac{\Delta[GM(a_7)]}{4\pi Ga_7^3}
\eeq
and the velocity dispersion (constant temperature) throughout the sphere
would be
\beq
\sigma^2 =\frac{\Delta[GM(a_7)]}{2a_7}
\eeq

We have investigated this procedure, but do not consider its
results meaningful.  Any assumed $DM$ interior to the orbit
of Jupiter is almost certainly fictional since gravitational
perturbations from Jupiter would eject it in a short time.
On the other hand the progression  from Uranus to Neptune implies
a best-fit
decreasing $M(r)$ (after subtracting out the masses of the
planets themselves) which is inconsistent with the assumption
of an isothermal distribution, or any other spherical mass distribution.

\subsection{Discussion}

There is debate as to the extent to which bodies of normal baryonic
matter formed at the time of formation of the sun, interior to the
orbit of Neptune, would be expected to survive.  Modern theories of
comets (see Bailey, Clube and Napier, 1990, for a review) are based on
formation of the Oort cloud by means of ejection of such bodies from
interior to the orbit of Neptune by the outer planets.  The efficacy of
such a mechanism was shown by Fernandez (1978).  It has been shown,
however, by Duncan, Quinn and Tremaine (1989) that stable circular
orbits are likely to  exist interior to Neptune (see, however, Gladman and
Duncan, 1990, Holman and Wisdom, 1993).  Thus our bound on the amount of normal
matter interior to Neptune's orbit may be applicable to models of Oort cloud
formation.

Our result in this paper -- any spherically symmetric distribution of
nonluminus matter must be less than a few times
$10^{-6}$ solar masses out to Neptune --  shows rather clearly that the
sun could not have captured all the dark matter the Bahcall analysis
requires   in the solar neighborhood
($0.1\;M_{\odot}$/pc$^3$  with ``neighborhood'' defined as within
 0.1 pc) into any distribution as centralized as those
considered here.  That is, the Bahcall  analysis says that the $DM$
density should be about equal to the density of luminous
matter,  but this much $DM$ about the sun captured during its
formation and retained past Saturn is inconsistent with our result.  In this
connection, note that, as pointed out by Tremaine (1991), tidal forces from
passing stars would not be effective in displacing dark matter
interior to the Oort belt at $10^4$ AU.  Our result may focus
the Bahcall dark matter problem by decreasing the
possibility of its being resolved by small bodies of normal
matter.  It argues for either: (1)  ``new particle physics'', e.g.
elementary particles that cannot radiate but can dissipate sufficiently
to condense in the galactic disk but not sufficiently to be captured by
the sun during its formation; or else (2) ``new astrophysics,'' e.g.
large numbers of brown dwarfs.

We consider now the question of how much DM the sun
could be expected to capture gravitationally during its
formation.  Conditions for capture during formation of the
sun of a weakly interacting particle must be
\beq
v^2/2<G/r\;(dM/dt)\Delta\,t;\qquad\qquad\qquad
v\Delta\, t< r\,.
\eeq
That is, to be captured a particle must be moving slowly
enough that it (a) does not leave the scene during formation
of the sun and (b) has a velocity less than the escape
velocity.  Taking from Shu et al. (1987) that half the mass
of the sun accumlates in $2.5 \times 10^5$ yr, one sees that
the sun would be expected to capture all dark matter within
0.1 pc moving slower than about 0.3 km/s.

Thus our result puts no constraint on dark matter that is
weakly interacting only, spread relatively uniformly over a
spherical galactic halo, and moving with a gaussian
distribution about the galactic virial velocity of 300 km/s.
This is because the halo density is expected to be about
$10^{-4}M_{\odot}$ /pc$^3$
 so the amount captured should be $
10^{-9-3-4}\, M_{\odot}\sim 10^{-16}\, M_{\odot}$.  While it is not
completely clear  that relaxation mechanisms cannot
enhance gravitationally the density of weakly interacting
DM particles in the galactic disk,  in the
Appendix  we present a calculation that makes such a
scenario highly doubtful.

We can provide one possible direction, beyond those
discussed in Section I above, in which particle models with dissipation
may be found (although whether nature chooses one of them is a very
different question).  If particle $X$ dissipates energy by scattering,
it should have a cross section $\sigma $ such that it will scatter and
release  some energy  at least once in a time $t$, on the order of $10^9$
years.  For scattering off protons, electrons, or Hydrogen we can
calculate the cross section needed for dissipation since we know the
proton density ($n$).  Assuming a virial  velocity, $v$, for
$X$-particles, we have from  $$ n\sigma vt_1=1 $$
with
\beqast
&& V_{gal}\sim 10^{70}\, cm^{3} \\
&& n_p \sim 0.01 \, cm^{-3} \\
&& v \sim 300 \, km/s
\eeqast
a cross section of
$$\sigma \sim 10^{-22}\, cm^2\,.  $$
Such a large cross section is ruled out of course.

Now consider the universe to be dominated by a very light abundant
particle,  for example the axion $(a)$ with a dissipation mechanism in
an interaction $a+a\rightarrow a+a+Y$, with $m_Y\ll m_a$.  If $m_a\sim
10^{-5}eV$ (see, for example, Kolb and Turner (1990)) and $\Omega_a\sim
1$ then

$$\rho_a=\rho_c\sim 10^3 eV\, cm^{-3}$$
which implies
$$ \overline{n}_a \sim 10^8\;cm^{-3} \,.$$
The axion number density in the galaxy could be as large as
$$ n_a\;{(\rm galatic)} \sim  \frac{\overline{n}_p\;({\rm galaxy})}
{\overline{n}_p\; ({\rm universe})}\; \overline{n}_a\,.$$
Since
$$\overline{n}_p\;({\rm universe})=0.01\, \frac{\rho_c}{m_p} = 10^{-8}\;
cm^{-3}$$
the number of axions in the galaxy is 10$^{16}$ larger than the number
of protons.  Thus a cross section 10$^{16}$ smaller than the
$10^{-22}\, cm^2$ above would give significant dissipation.  A zero mass
Majoran would be a candidate for $Y$.  In short, one direction for $DM$
models with dissipation is that of $\Omega$ dominated by a very light,
and hence very abundant, but non-relativistic, particle with significant
inelastic scattering.

Finally, we note that our analysis can be extended to address the
question of the existence of a belt of cometary matter in the region
just past the orbit of Neptune. As noted, such a belt has been
postulated by a number of authors, Kuiper (1951), Duncan et al. (1988),
in order to explain the high relative frequency of short period comets.
Interest in the possibility of such a belt has increased recently with
the observation of   candidate objects by Jewett and Luu (1993).  The
techniques of the present work, generalized to mass distributions that
are not spherically symmetric, should be able to place limits on the
mass and location of such a belt or to detect its presence.  Such an
effort is under way.  In the meantime, a gross estimate of a bound can
be made by equating the approximate attraction of such a belt on Neptune
to the attraction of a spherically symmetric density of DM sufficient
to saturate our present bound on DM interior to  Neptune's orbit.  In
the approximation that the distance $r_{BN}$ of Neptune to the belt is
much smaller than $r_N$ the semi-major axis of orbit of Neptune, we
have, for a belt 10 AU past Neptune
$$ F\sim 2G/r_{BN}\times M_B/2\pi\; (r_{BN}+ r_N) \leq G\Delta M/r_N^2 $$
or
$$ M_B \leq \pi\Delta Mr_{BN}\, (r_{BN}+r_N)/r_N^2\sim \Delta M \,.$$
In this crude approximation, a belt 10 AU past Neptune must be less than
a few Earth masses. However this approximate calculation does not take into
account more sensitive effects of such a belt such as precession of the line
of nodes of Neptune.  (See, for example, Whipple, 1964.)

In summary, we recapitulate the principal results of this paper and the
earlier one, Anderson et al. (1989), in Table 2.  We note that we now
have a limit on the amount of dark matter in orbit about the sun in a
spherically symmetric distribution interior to Neptune of less than an
earth mass and a Uranus limit of  about 1/6 of an earth mass.

\setcounter{equation}{0}
\renewcommand{\theequation}{A.\arabic{equation}}

\subsection*{Appendix:~~Gravitational Scattering and Disk \newline ~~~~~~~~~
Dark Matter
Density}

We investigate here whether the density in the galactic disk of
weakly-interacting DM might be enhanced over the density in the
galactic halo.  The mechanism in question would be that of repeated soft
gravitational scattering of galactic halo dark matter particles off giant
molecular cloud complexes.  This would be essentially the inverse of the
mechanism of Spitzer and Schwarzschild (1951) by which scattering off
giant molecular cloud complexes explains the greater velocity dispersion
of older stars.  We show this does not work, a result that may be
intuitive from thermodynamics.

We approximate the galaxy as a slab of clouds of mass $m_c$, with density
$n_c$, traveling with  constant velocity $v$.  We find the effect of this
distribution of clouds
\beq
f_c(v_c) = n_c\delta (\vec{v}_c - \vec{v}_1)
\eeq
on an initial gaussian distribution (in velocity space) of halo DM
particles
\beq
f(v) = A\, e^{-v^2/v_2^2}\,.
\eeq
We calculate within the local approximation to the ``master equation''
as formulated in Binney and Tremaine (1987).  We have then
\beqra
df(v)/dt =\Gamma(f) &=& -\sum_1^3 \, \frac{\partial}{\partial
v_i}\;\left[f(v)D(\Delta v_i)\right] \nonumber \\
& & +\haf\; \frac{\partial^2}{\partial v_i\partial v_j}\;
\left[f(v)D(\Delta v_i\Delta v_j)\right]
\eeqra
where Appendix 8.A of Binney and Tremaine (1987) gives
\beqra
D(\Delta v_i) &=& - 4\pi G^2m^2_a\ln\La\int\, \frac{f_c(v_c)}{v_0^3}\;
v_{oi}d^3v_c \\
D(\Delta v_i,\Delta v_j) &=& +4\pi G^2m^2_a\ln\La\int\,
\frac{f_c(v_c)}{v_0^3}\;\left(\delta_{ij}-
\frac{v_{oi}-v_{oj}}{v_0^2}\right) \, d^3v_c\,.
\eeqra
Here $v_0=v-v_c$ and the ``Coulomb logarithm,'' ln$\Lambda$, is of the order
ln ($R_{\rm Gal}/R_{{\rm cloud}})$.  Substituting (A.1,2,4,5) into (A.3)
gives
\beqra
d\;f(v)/dt &=& K\left\{\partial_i\left[e^{-v^2/v_2^2}\,
\frac{(v-v_1)_i}{|v-v_1|^3}
\right] \right. \nonumber \\
&+& \left.\haf \partial_i\partial_j\; e^{-v^2/v_2^2}\,\left[ \frac{\delta_{ij}}
{|v-v_1|} - \frac{(v-v_1)_i(v-v_1)_j}{|v-v_1|^3}\right]\right\}
\eeqra
where $K=4\pi G^2m^2_c \,(\ln\La)\, n_c$.  Performing the differentiations in
(A.6) gives
\beqra
df(v)/dt &=& Ke^{-v^2/v_2^2}\left\{ \left[ \,\frac{-2(v^2-v_1\cdot
v_2)}{v_2{}^2 |v-v_1|^3} + 4\pi\delta (\vec v-\vec v_1)\right] \right.
\nonumber \\ && +\frac{2}{v_2^4}\; \left[ \frac{v^2}{|v-v_1|} -
\frac{(v^2-\vec v\cdot \vec v_1)^2}{| v- v_1|^2}\right] \nonumber \\
&& + \frac{1}{v^2}\, \left[\frac{2(v^2-\vec v\cdot\vec v_1)}{|v-v_1|^3}
- \frac {3}{|v-v_1|} + \frac{(2\vec v-\vec v_1)\cdot (\vec v -\vec v_1
)}{|v-v_1|^3}\right]
\nonumber \\
&& -4\pi\delta (\vec v-\vec v_1)\Bigg\}\,.
\eeqra
The first square bracket comes from the first term in (A.6).  (A.7)
becomes
\beq
df(v)/dt =\frac{ke^{-v^2/v_2^2}}{v_2^4|v-v_1|^3}\,
\left\{ 2\left[v^2v_1^2-(\vec v\cdot \vec v_1)^2\right] + v_2^2 \left[
3\vec v\cdot \vec v_1 -2v_1^2-v^2 \right]\right\}\,.
\eeq
Letting $v=v_1+\eta$ in (A.8) gives
\beq
df(v)/dt = \frac{ke^{-v^2/v_2^2}}{\eta^3\,v_2^4}\, \Big[
\!\vec{v}_2^2(\vec{v}_1 \cdot \vec\eta) - (2v_1^2 - v_2^2)\eta^2-(\vec
v_1\cdot \vec\eta)^2 \Big]\,. \eeq

Equation (A.9) is our principal result.  We see that the $\eta^{-3}$ factor
provides an enhancement to the rate for scattering of $DM$ particles off
molecular clouds when these particles have small velocitities relative to
the clouds.  However the first term in brackets in (A.9) merely removes
$DM$ particles with velocities somewhat less than $v_1$, and adds $DM$
particles with velocities somewhat greater, with no net difference in
total density.  The other terms in (A.9) are negative ($2v^2_1$ being
greater than $v_2^2$).  Thus the net effect of (A.9) is to remove $DM$
particles from the galactic disk by scattering off clouds, not to add to
the density of disk $DM$ .  Such a result might be expected
on the basis of general principles of statistical mechanics: increasing
the density in the two-dimensional disk corresponds to decreasing the
entropy of the three-dimensional $DM$ system.

\pagebreak

\centerline{\bf Table Captions}

\begin{description}
\item[Table 1.]  {\bf Range Points to Jupiter, Uranus and Neptune.}\\
We collect the range points to Jupiter, Uranus and Neptune.  The
analysis is described in Standish (1990).  The Voyager points for
Jupiter and Neptune have not been previously published.

\vspace{12pt}
\item[Table 2.] {\bf  Limits on Dark Matter.} \\
Line 1 reproduces the results of Anderson et al. (1989); line 2
shows the improvement from the new fitting technique without the Neptune
ranging point. Lines 3-5 use the new, improved fitting procedure as described
in the text and ranging points as indicated.  The time argument is the Julian
date JD associated with the JPL ephemerides (the relativistic coordinate time
referenced to the solar-system barycenter).

\end{description}
\pagebreak

\vspace*{24pt}
\centerline{\bf Table 1}

\vspace{12pt}
\begin{center}
\begin{tabular}{llr@{$\pm$}lr@{$\pm$}l}
{}~~~~~Spacecraft &  Date (JD) &
\multicolumn{2}{c}~Geometric Distance~~ &\multicolumn{2}{c}~ DE200 Residual\\
& &~~ (1-way Km) & &~~(1-way Km)
 \\ \\
Jupiter Pioneer 10 & 2442020.50 & 825852471.1~~&~12 &~~~~~~$-$5.6&12
\\[6pt]
 Jupiter Pioneer 11 & 2442384.50 & 731437233.5~~&~3     &6.0&3\\[6pt]
Jupiter Voyager 1 & 2443938.00 &678931390.1~~&~4      &114.1&4\\[6pt]
Jupiter Voyager 2 & 2444064.50 &932054679.9~~&~4     & 96.1&4\\[6pt]
Uranus  Voyager 2 & 2446455.25 & 2965361517.0~~&~1       &147.3&1\\[6pt]
Neptune Voyager 2~~~ & 2447763.67~~~&  4425522117.1~~&~1 &8224.0&1
\end{tabular}
\end{center}

\pagebreak

\vspace*{24pt}
\centerline{\bf Table 2}

\vspace{12pt}
\begin{center}
\begin{tabular}{llll}
\multicolumn{4}{r}  {Limits on Dark Matter (in units of
$10^{-6}M_{\odot}$)~}\\
\\
Spherically symmetric & Uranus & ~~Neptune & ~~Jupiter \\
{}~~~~~~Distribution &&& \\
\\
Anderson et al.  (1989)  &$~~0.4\,\pm 2.8\,$ &  ~$-0.4\pm 114$  & \\
Uranus ranging  & & & \\
\\
Uranus ranging with & $0.32\, \pm 0.49$ & ~~~$~38\,\pm 108$ & \\
all planets refit & & & \\
\\
Uranus, Neptune ranging & $0.32\pm 0.49 $ & ~$-1.9\pm \,1.8$ & \\
with all planets refit & & & \\
\\
Uranus, Neptune, &$0.33\pm0.49$ & ~$- 1.9 \pm \,1.8$ & ~$0.12 \pm
0.027 $  \\  Jupiter ranging~~ & & & \\
\\
Including Jupiter ranging \\
with $M_{\odot}$ for Jupiter &$0.26 \pm 0.49$ &~ $- 2.0\pm1.8\, $ &\\
fixed by inner planets

\end{tabular}
\end{center}
\pagebreak

\section*{Acknowledgements}

It is a pleasure to acknowledge  stimulating conversations with
Professor E.W. Kolb, as well as helpful interchanges some years ago with
Professor David Spergel and valuable correspondence with Professor S.
Tremaine on the appendix. We have benefited from several communications
with E. Myles Standish, Jr. We would also like again to thank Dr.
Jeffrey Briggs for bringing together the JPL and non-JPL authors in 1987.
J.D.A. and E.L.L. acknowledge support from the Pioneer Project Office,
NASA Ames Research Center, under Letter of Agreement ARC/PP017; and
support from the Ultraviolet/Visible/Gravitation Astrophysics Office
of the NASA Astrophysics Division.  Their work was preformed at the Jet
Propulsion Laboratory, California Institute of Technology, under
contract with NASA.  The work of DAD was supported in part by the U.S.
Department of Energy under Grant No. DE-FG03-93ER40757.  The SMU work
was supported in part by the Texas National Research Laboratory
Commission.

\section *{References}

\be

\item[{}] Alcock, C., et al., 1993, \underline{ Nature}, {\bf 365}, 621.

\item[{}] Anderson, J.D., Lau, L.L., Taylor, A.H., Dicus, D.A., Teplitz,
D.C., and Teplitz, V.L., 1989, Ap. J. {\bf 342}, 539.

\item[{}] Auborg, E., et al., 1993, Nature {\bf 365}, 623.

\item[{}] Backman, , and Gillett, 1987, in {\it Cool Stars, Stellar
Systems, and the Sun} (ed. Linsky, J.L. and Stencil, R.E.) 340
(Springer, Berlin).

\item[{}]  Bahcall, J.N., 1984, Ap. J. {\bf 287}, 926.
\item[{}] Bahcall, J.N., 1992, Ap. J. {\bf 389}, 234.

\item[{}] Bailey, M.E., Clube, S.V.M., and Napier, W.M., 1990, \ul{The
Origin of Comets}, Pergamon Press, N.Y.

\item[{}] Binney, J., and
Tremaine, S., 1987, \ul{Galactic Dynamics},  Princeton University Press,
Princeton, New Jersey.

\item[{}] Caldwell, D.O., 1990, Phys. Rev. Lett. {\bf 65}, 1305.

\item[{}] Duncan, M., Quinn, J., and Tremaine, S., 1988, Ap. J.  {\bf
328}, 269.
\item[{}] Duncan, M., Quinn, J., and Tremaine, S., 1989, Icarus {\bf
82}, 402.

\item[{}] Duncan, M., Quinn, J., and Tremaine, S., 1990, Ap.
J. {\bf 355}, 667.

\item[{}] Faulkner, J., and Gilliland, R.L., 1985, Ap. J. {\bf 299},
594.

\item[{}] Fernandez, J.A., 1978, Icarus {\bf 34}, 173.

\item[{}] Gladman, B., and Duncan, M., 1990, Astron. Jour. {\bf 100}, 1680.

\item[{}] Gross, D.J., et al., 1985, Phys. Rev. Lett. {\bf 54}, 502.

\item[{}] Guth, A., 1981, Phys. Rev. {\bf D23}, 342.

\item[{}] Hodges, H.M., 1993, Phys. Rev. {\bf D47}, 456.

\item[{}] Hoffman, G.L., et al, 1993, Astron. Jour. {\bf 106}, 39, and
references  therein.

\item[{}] Holman, J.J., and Wisdom, J., 1993, Astron. Jour. {\bf 105}, 1987.

\item[{}] Jewett, D., and Luu, J., 1993, {\it Nature} {\bf 362}, 730;
and IAU Circulass 5730, 5865, 5867.

\item[{}] Kane, G., 1993, \underline{Modern Elementary Particle Physics},
Addison
Wesley Publishing Co., N.Y.

\item[{}] Khloper, M.Y., et al., 1991, Astron., Zh {\bf 68}, 45.

\item[{}] Kolb, E.W., and Turner, M.S., 1990, \ul{The Early Universe},
Addison-Wesley Publishing Co., New York.

\item[{}] Kuiper, G.P., 1951, in {\it Astrophysics} (ed. Hynek, J.A.)
357-424 (McGraw Hill, N.Y.)

\item[{}]Mulchaey, J.S., Davis, D.S., Mushotzky, R.F., and Burstein, D., 1993,
Ap. J. in press.

\item[{}] Oort, J., 1950, Bull. Astron, Inst. Meth. {\bf 11}, 91.

\item[{}] Press, W.H., Teukolsky, S.A., Vetterling, W.T., and Flannery, B.P.,
1992, {\it Numerical Recipies}, Cambridge University Press, Cambridge, England.

\item[{}] Press, W.H., and Spergel, D.N., 1985, Ap. J. {\bf 296}, 679.

\item[{}] Rubin, , et al., 1985, Ap. J. {\bf 289}, 81.

\item[{}] Shu, F.H., Adams, F.C., and Lizano, S., 1987, Ann. Rev. Astron.
Astrophys. {\bf 25}, 23.

\item[{}] Smoot, G.F., et al., 1992, Ap. J. {\bf 396}, L1.

\item[{}] Spitzer, L.S., Jr.,  and Schwarzschild, M., 1951, Ap. J. {\bf
114}, 385.

\item[{}] Standish, E.M., Jr., 1990, {\it The observational Basis for
JPL's DE200, the Planetary Emphemerides of the Astronomical Almanac.}
Astron. and Astrophys. {\bf 233}  252-271.

\item[{}] Standish, E.M., Jr.,  Newhall, X. X., Williams, J.G., and
Yeomans, D.K., 1992,  {\it Orbital Ephemerides of the Sun, Moon, and
Planets.}  In {\it Explanatory Supplement to the Astronomical Almanac},
ed. P.K. Seidelmann. 279-323. Mill Valley: University Science Books.

\item[{}] Standish, E.M., Jr., 1993, Astronom. J. {\bf 105}, 2000.

\item[{}] Starkman, G.V. et al., 1990, Phys. Rev. {\bf D41}, 3594.

\item[{}] Tremaine, S., 1990, in \underline{Baryonic Dark Matter}, D.
Linden-Bell and G. Gilmore, ed; page 37, Klewer
Academic Publishers, Boston.

\item[{}] Turner, M.S., 1991, Physica Scripta {\bf T36}, 167.

\item[{}]  Udalski, A., et al., 1993, ACTA Astronomica {\bf 43}, 289.

\item[{}]  Weissman, P.R., 1990, Nature {\bf 344}, 825.

\item[{}]  Whipple, F.L., 1964, Proceeding of the National Academy of
Sciences {\bf 51}, 711.

\ee

\end{document}